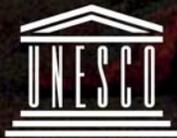

THE RIGHT TO
DARK SKIES

EL DERECHO
A LOS CIELOS
OSCUROS

United Nations
Educational, Scientific and
Cultural Organization

Organización
de las Naciones Unidas
para la Educación,
la Ciencia y la Cultura

# Protecting the Dark Skies of Chile: Initiatives, Education and Coordination

**Guillermo Blanc**
*Universidad de Chile / Sociedad Chilena de Astronomía (SOCHIAS), Chile*

**Abstract**

During the next decade, Chile will consolidate its place as the 'World Capital of Astronomy'. By 2025, more than 70% of the world's infrastructure for conducting professional astronomical observations will be installed in the Atacama Desert in the north of the country. The amazing scientific discoveries these telescopes produce have a direct impact on our understanding of the cosmos, and protecting this 'window to the universe' is fundamental in order to ensure humanity's right to contemplate the night sky and decipher our origins.

As a country, Chile faces the challenge of fighting light pollution and protecting its dark skies in a context of sprawling urban growth and an ever-expanding mining industry that shares the same territory with astronomical observatories. The Chilean Astronomical Society (Sociedad Chilena de Astronomia, SOCHIAS) plays an active role in protecting dark skies through a series of initiatives involving educational programmes, aiding in the development and enforcement of public policy and regulation, and seeking the declaration of Chile's best astronomical sites as protected heritage areas, both at the national and international levels.

Whilst describing our experiences, I highlight the importance of approaching the problem of light pollution from all sides, involving all the relevant actors (communities, national and local governments, lighting industry, environmentalists, astronomers and others). I also discuss how communication and timely coordination with potential problematic actors (like industries, cities and some government agencies) can be an effective tool to transform potential enemies into allies in the fight for the protection of the night sky.

**Resumen**

Durante la próxima década, Chile afianzará su lugar como "Capital Mundial de la Astronomía". Para el 2025, más del 70% de la infraestructura mundial para realizar observaciones astronómicas profesionales estará instalada en el Desierto de Atacama, al norte del país. Los asombrosos descubrimientos científicos que generan estos telescopios tienen un impacto directo en nuestro entendimiento del cosmos, y proteger esta "ventana al universo" es fundamental para garantizar el derecho de la humanidad a contemplar el cielo nocturno y descifrar nuestros orígenes.

Como país, Chile enfrenta el desafío de luchar contra la contaminación lumínica y proteger sus cielos oscuros en un contexto de crecimiento urbano descontrolado y con una industria minera en continua expansión que comparte el mismo territorio que los observatorios astronómicos. La Sociedad Chilena de Astronomía (SOCHIAS) desempeña un activo papel a la hora de proteger los cielos oscuros a través de una serie de iniciativas que implican programas educativos, ayudando en el desarrollo y aplicación de políticas públicas y regulaciones, y solicitando la declaración de los mejores sitios astronómicos de Chile como áreas de patrimonio protegidas, tanto a nivel nacional como internacional.

Al tiempo que describo nuestras experiencias, subrayo la importancia de abordar el problema de la contaminación lumínica desde todos los ángulos, implicando a todos los actores relevantes (comunidades, gobiernos nacionales y locales, industria de la iluminación, ecologistas, astrónomos y otros). Analizo también cómo la comunicación y la coordinación oportuna con potenciales actores problemáticos (como industrias, ciudades y algunas agencias gubernamentales) puede ser una eficaz herramienta para transformar a potenciales enemigos en aliados en la lucha por la protección del cielo nocturno.

**Chile: World Capital of Astronomy**

During the first half of the twentieth century, the largest telescopes in the world, like the 100-inch Hooker Telescope at Mount Wilson and the Palomar 200-inch Hale Telescope, were mostly concentrated in southern California, in the continental United States. In the 1950s and 1960s, American and European astronomers started scouting potential sites in the Atacama Desert in northern Chile. This was part of an effort to find places offering superb observation conditions and also ensuring low levels of light pollution in the future. The urban sprawl of cities like Los Angeles and San Diego was quickly diminishing the capabilities of the most powerful telescopes in the world, and the low population density of the Atacama Desert and its remoteness seemed to offer decades, even centuries, of dark skies ahead.

With the construction of the Cerro Tololo Inter-American Observatory (CTIO) in 1963 by the United States National Optical Astronomical Observatory (NOAO), the establishment of the La Silla Observatory in 1965, operated by the European Southern Observatories (ESO), and the start of operations at the Las Campanas Observatory in 1969, built by the Carnegie Institution for Science, the Atacama Desert started its journey towards becoming the most important hub of astronomical observing facilities in the world.

Today, the Atacama hosts about 30% of the total optical telescope collecting area in the world, including flagship projects like the four 8.2 metre ESO Very Large Telescopes (VLT), the two 6.5 metre Magellan Telescopes at Las Campanas and the 8.1 metre Gemini South Telescope. Chile's share of the world's optical astronomical infrastructure will increase to about 70% with the completion of projects currently in construction, such as the Large Synoptic Survey Telescope (LSTT), the 25 m Giant Magellan Telescope (GMT) and the 39 m European Extremely Large Telescope (E-ELT).

Furthermore, considering telescopes that observe at radio wavelengths, where light pollution can be present in the form of radio interference from communications and other sources, Chile already hosts most of the world's radio telescope collecting area, primarily thanks to the Atacama Large Millimetre/sub-millimetre Array (ALMA). This is the largest and most powerful radio telescope in the world, consisting of a sixty-six 12 m antenna interferometer.

All this investment in astronomical infrastructure has undoubtedly transformed the Atacama Desert into the 'World Capital of Astronomy' and has brought enormous benefits to the country on several fronts. Chile and its government see the development of astronomy as an opportunity for national development. The scientific cooperation between international observatories and Chilean universities has translated into enormous advances in the quality, quantity and impact of astronomy, astrophysics and astronomical instrumentation research that is conducted in Chile. Today, Chilean universities are at the forefront of astronomical research and compete side by side with universities in Europe and North America.

The international observatories in the Atacama have also sparked the attraction and training of highly qualified professionals in several fields of engineering, physical and mathematical sciences and computer science. The transfer of knowledge, expertise and technology between astronomical observatories, universities and the private sector offers enormous opportunities for the development of advanced technology engineering companies in Chile, an aspect of strategic importance for future economic growth and international competitiveness. Furthermore, other areas like STEM education, the growing industry of astrotourism and the protection of the country's natural and cultural heritage have all benefited in one way or another from the development of astronomy in the country.

**Light Pollution in Chile**

Until the early 1990s, light pollution around important astronomical observatories in Chile was minimal. Most sites, even those close to large cities, like Cerro Tololo near La Serena, showed measured sky brightness consistent with natural (virgin) levels. During the 1990s, Chile saw enormous economic and population growth, with a sustained annual GDP increase ranging from 5–10% and a 15% growth in population. This development continued over the next two decades and a significant fraction of it has happened in and around the Atacama Desert, the country's main mining region.

Figure 1 presents maps of the Atacama's sky brightness levels at the zenith computed using night time satellite imagery and atmospheric light propagation models. It is clear that the areas affected by light pollution have increased substantially over the years. Already in the mid-1990s, sites close to cities like Cerro Tololo and Cerro Pachón showed small but significantly measured enhancements in sky brightness above the natural levels. Today, the levels of light pollution at Tololo and Pachón are increasing alarmingly and remote sites that previously showed virgin dark skies, such as Las Campanas and La Silla, now sit at the edge of regions significantly affected by light pollution.

The problem of light pollution is now evident to the naked eye in remote sites such as Las Campanas. Figure 2 shows a long exposure all sky image of the night sky at this site. The twin 6.5 metre Magellan telescopes are seen at the bottom. On the horizon, one can clearly identify the glowing sky above cities such as La Serena, Coquimbo, Vallenar and Copiapó. One can also see significant sources of light pollution from nearby mining operations, and the lights illuminating crossroads and toll plazas on a nearby stretch of the Panamerican Highway.

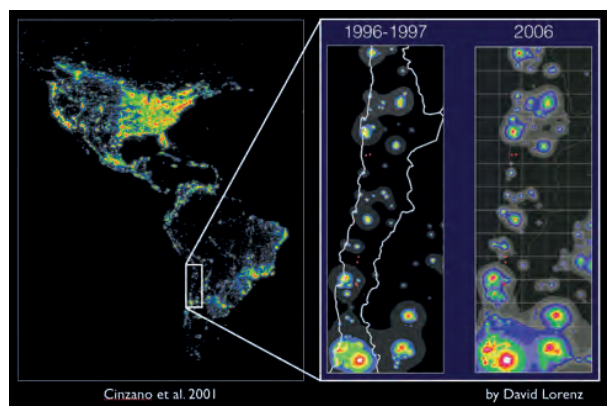

Figure 1. Evolution of light pollution in northern Chile 1996-2006. Red squares mark the positions of the main optical astronomical observatories in Chile. From south to north: Pachón and Tololo, La Silla, Las Campanas, Paranal and Armazones. © Guillermo Blanc adapted maps by P. Cinzano, F. Falchi, University of Padova; C. D. Elvidge, NOAA National Geophysical Data Center, Boulder; and David J. Lorenz, University of Wisconsin-Madison.

A few decades ago, these levels of light pollution at such remote sites were unimaginable. Astronomers, in pursuit of good image quality and low sky brightness, usually conduct their observations by pointing their telescopes at high elevations above the horizon. Astronomical observations, therefore, tend to avoid the areas most dramatically affected by light pollution. If the sustained increase in sky brightness is not controlled appropriately, the sky brightness of the entire night sky can be seriously affected, even towards the zenith, forever degrading the 'windows to the universe' that the dark skies above the Atacama Desert provide us.

Beyond the direct detrimental effects of light pollution on the quality of astronomical observations, several other negative effects are now recognized to be associated with excessive, unnecessary levels of illumination. These include, among other things, distortions in people's sleep patterns and associated health problems, disruption of migration patterns and the hunting and reproductive behaviour of certain animal species and negative environmental and economic impacts associated with the unnecessary overproduction of electric power that is wasted on illuminating the sky. Moreover, in the particular case of Chile, light pollution threatens to take away all the evident benefits to our society that the development of astronomy is currently providing and will provide in the future.

**Light Pollution Regulation in Chile**

Astronomers in Chile had already began expressing serious concerns about the problem of light pollution and its dramatic evolution in the 1990s. Their concerns were well received by the Chilean government and prompted the first regulation on light emissions in the country's history. The 1998 Supreme Decree N° 686 of the Ministry of Economy regulated the characteristics of exterior light fixtures in the three northern regions of the country (Antofagasta,

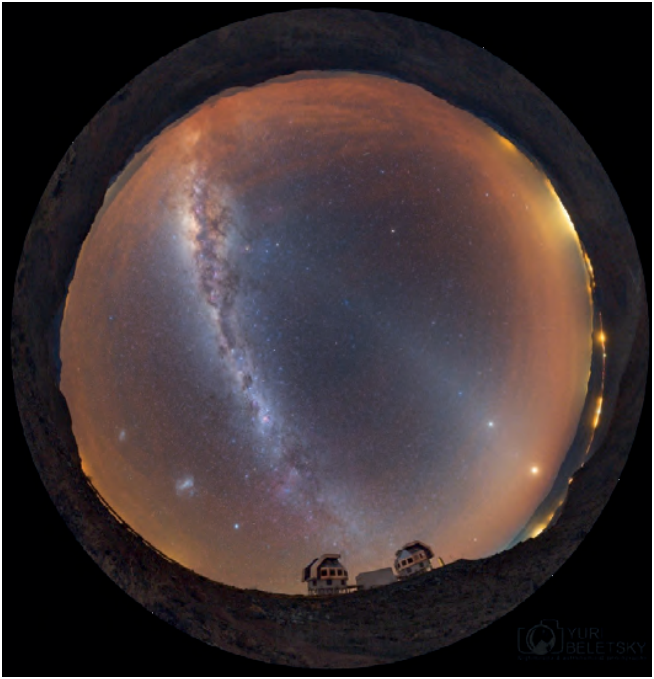

Figure 2: Present day levels of light pollution at Las Campanas Observatory. The cities of La Serena, Vallenar and Copiapó, as well as lightning on the Panamerican Highway and some nearby mining operations are clearly visible on the horizon. © Yuri Beletsky, Las Campanas Observatory.

Atacama and Coquimbo), where most of the observatories are located. Before the 1998 decree, the only protection astronomical observatories received was the designation of 'Scientific Interest Zones for Mining Purposes' outlined in the Chilean Mining Code, which defines relatively small buffer zones where no mining activities can be performed without the expressed authorization by the President of the Republic.

Decree Nº 686 was followed by the creation of the Office for Protection of Sky Quality in Northern Chile (OPCC in Spanish) in May 2000. The OPCC currently operates under an agreement between the Chilean government, a consortium of international observatories installed in the country and more recently the Chilean Astronomical Society (SOCHIAS in Spanish). This office, under the leadership of Director Pedro Sanhueza, has been the most important actor in the fight against light pollution in Chile during the last two decades. A 10-year effort led by the OPCC and others translated into widespread compliance with the light emissions regulation established in Decree Nº 686. This enormously mitigated the growth of light pollution during the 2000s, which, given the growth of cities in the region, could have been much worst during that decade.

In the early 2010s, the advent of new illumination technologies, mainly the widespread availability of energy efficient Light Emitting Diodes (LEDs), with a strong emission component at blue wavelengths[1] and the popularisation of bright large area LED screens for use in advertisement, made the Decree Nº 686 insufficient in terms of providing adequate protection of dark skies in the north of Chile. This prompted the development of new regulation, now coming from the Ministry of Environment.

The new Supreme Decree Nº 43 of the Ministry of Environment officially replaced the original Decree Nº 686 in 2014, imposing much stricter regulations concerning the characteristics of lighting fixtures and illuminated signs used in outdoor lighting, advertisement, industrial illumination (including mining) and sports events. Of particular importance is the requirement for all outdoor lighting fixtures to have 'full cut off' (that is, zero emissions above the horizontal), strict limits on the fraction of the total emissions that can be output at blue frequencies and the establishment of maximum levels of illumination at floor level. While this new decree has the potential to be a powerful tool for controlling light pollution in the Atacama Desert, its implementation has suffered delays. Ensuring a significant level of compliance will require major efforts in education and coordination between the different actors involved in the problem of light pollution. It is here where SOCHIAS, in partnership with the OPCC, is playing a central role in promoting the protection of dark skies.

### The Work of SOCHIAS and the OPCC

SOCHIAS is a scientific society that brings together most of the professional astronomers working in Chile. It was created in the year 2000 and it currently has more than 160 members, including researchers, professors and students at Chilean universities, as well as astronomers and engineers working at different astronomical observatories in the country. Since 2015, SOCHIAS has established a close partnership with the OPCC, the international observatories and the Chilean government, with the goal of fighting light pollution and protecting Chile's dark skies.

---

1. Blue light produces more light pollution than red light as it scatters more easily in the atmosphere.

From the perspective of SOCHIAS, we see the problem of light pollution as one involving many diverse actors, who play different roles. These include, on one side, the astronomers, the environmentalists and the regulatory government agencies like the Ministry of Environment that have a vested interest in the fight against light pollution. On the other side, local communities, city governments, private industries and other national and regional government agencies related to energy and transportation, can all see different incentives to either produce more light pollution or reduce its levels. For example, a local city government might want to increase the level of lighting in its streets in an effort to reduce night-time crime in the area, or it might want to reduce its electric bill by adopting better illumination practices that reduce the amount of wasted energy spent on illuminating the sky. If the problem of light pollution is to be addressed effectively, we must involve all the different actors, understand their needs and priorities, and find common ground and synergies that reasonably protect their interests while also protecting our dark skies.

In this context, SOCHIAS is involved in several initiatives aimed at protecting the dark skies of northern Chile, which target different actors in the problem of light pollution. These initiatives fall under two main categories: education initiatives and coordination initiatives. Both are essential tools in our fight to protect the unique resource that our country has in its night sky.

**Education Initiatives**

Education starts from raising awareness about the existence of the problem of light pollution and the importance of the protection of dark skies. In terms of strategy, it is fundamental to understand that light pollution is a 'problem' that, at least in a country like Chile, is seldom acknowledged as such. Most people are either not aware of the existence of light pollution in the first place, or consider it to be a minor issue that should be given low priority when it comes down to enacting and enforcing environmental protections. Therefore, the starting point for our work on light pollution education is the general public, and the most effective tool to reach this public is the press. In SOCHIAS, we try to maintain a constant presence of news related to the protection of dark skies in the national press. During 2015, we were mentioned approximately twelve times in the national newspapers, television and radio shows. We also encourage all the members of our society to mention the subject of light pollution whenever they are interviewed about their scientific discoveries. This requires maintaining good relations with scientific journalists at the main news outlets in the country and also being fully available whenever an opportunity to raise the subject in the media arises.

Beyond the general public, a fundamental part of our strategy is to target specific interest groups such as companies that fabricate, import and distribute hardware for public and industrial lightning. For this, in close collaboration with the OPCC, we organized two workshops during 2015 that directly targeted the lighting industry. In these workshops, one of which included a presentation by the Chilean Minister of the Environment, Pablo Badenier, and other government authorities, we exposed the subject of light pollution and the technical requirements that the law mandates for illumination products to be certified for installation in protected regions.

We are seeing a very positive response from these companies and have taken away an important lesson from this experience. Companies in the illumination business have an obvious interest in selling more of their products and therefore have traditionally been seen as adversaries of groups trying to decrease the levels of light pollution. As a result of the new relationship that has been established through these workshops, we are happy to see some companies embracing the fight against light pollution and viewing it as an opportunity to better market their products as environmentally friendly and up to standard with the strict requirements of the law. They also see the need for massive replacement of old non-compliant luminaires as a potential business opportunity. This is a textbook case of how communication, education and an interest in understanding the necessities of all parties involved in a problem can turn potential adversaries into allies.

Similar experiences are regularly seen with companies in the mining sector, who because of worker safety concerns have a strong incentive for maintaining high levels of illumination in their facilities. The experience of the OPCC shows how receptive these companies are when, by means of meetings and other instances of communication, they realise that by minimal investments they can maintain the same levels of safety while dramatically reducing their impact on light pollution. In the case of the mining sector, there is a strong incentive to reduce light pollution in order to offset other environmental impacts associated with mining and to maintain a good image of social responsibility and sustainability.

Maybe the most important actors in the protection of our dark skies are local city or county governments. Ultimately, these are the decision makers responsible for approving and executing most of the outdoor illumination projects in a particular region. Given the administrative structure of Chile, 'municipalities' are the fundamental government unit that we must have on our side in the fight against light pollution. To ensure this, SOCHIAS and the OPCC have started a full scale training programme that we are carrying out in all municipalities of the three regions of the country currently under the protection of the new light emission regulation (the Decree Nº 43 mentioned previously).

While municipalities have a lot of influence on the way illumination is organized and implemented in their jurisdictions, they very often lack the economic resources to update their illumination systems, and they depend a lot on the national and regional governments to provide the necessary funding for such projects. Also, municipalities are often not well informed about the requirements in terms of design and certification that lighting fixtures must fulfil, and this can translate into erroneous purchases and installations of non-compliant and pollutant illumination systems.

In order to alleviate these problems, during 2015 SOCHIAS and the OPCC secured a grant from the ESO-Chile Comité Mixto Fund, which is allowing us to allowing us to finance the above mentioned training programme for municipal staff across the Antofagasta, Atacama and Coquimbo regions. The first of these training programmes took place in June of 2016 in the cities of Coquimbo and La Serena. This programme includes a general course on 'Sustainable Illumination Practices' given by an expert on illumination design, in addition to two other sessions. The first is aimed at understanding the negative impacts of light pollution and the second is aimed at understanding the technical requirements that Chilean law mandates in terms of characteristics and certifications for new luminaires.

A fundamental goal of these instances is to also include regional authorities and local representatives of the national government, as well as members of the local communities (for example, interested parties like astrotourism business owners). In this way, we can start creating a network of support for the municipalities that can help them push forward the necessary projects to improve their outdoor illumination systems.

Finally, at SOCHIAS we have the responsibility of involving our members in the fight against light pollution. Surprisingly, most professional astronomers in the world are not actively involved in the protection of dark skies. Since the ability to observe the universe from places like the Atacama Desert is a fundamental part of their job, one would expect a much higher level of involvement in dark skies protection than what is currently seen. The low involvement of professional astronomers is caused in part by the enormously competitive and absorbing nature of academic work and research, but we have come to realise that there is another important factor: lack of information and opportunities for involvement. Almost all astronomers are aware of the problem of light pollution but do not know what they can do about it on an individual level.

Therefore, as a society we are trying to actively involve the Chilean astronomical community in our dark sky protection initiatives. Since 2015, we have reserved a space in the SOCHIAS Annual Meeting (the most important and highly attended scientific conference on astronomy in Chile) to have talks and open discussion about light pollution. We educate our own astronomers on what tools and resources they have at their disposal to play a leading role in the fight against light pollution. This includes providing them with material that they can use in their classes and during outreach talks, showing them how to use the existing systems to denounce violations of the legislation that regulates light pollution in the country and offering them opportunities for involvement in the different initiatives we lead. It is fundamental for us to actively involve the members of our society in the initiatives and activities that we push forward.

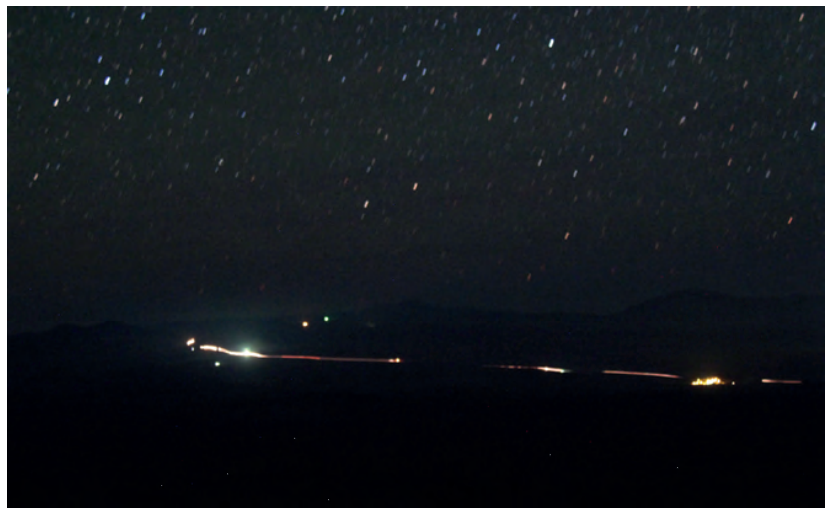

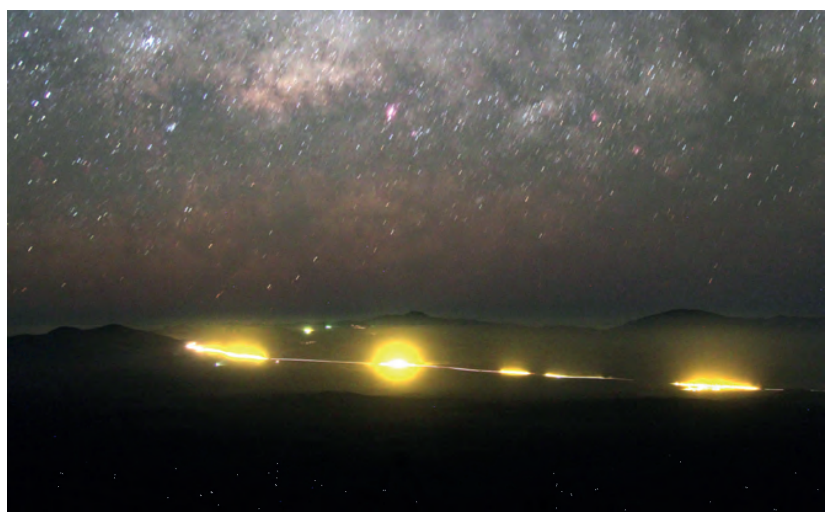

Figure 3: A stretch of the Panamerican Highway that crosses in front of the La Silla and Las Campanas observatories. The photo shows the highway before and after the installation of new illumination systems at intersections and toll plazas that are not compliant with the new regulatory standards for light pollution in the region. © Yuri Beletsky, Las Campanas Observatory.

## Coordination

While education is a fundamental tool in ensuring the long-term support of all the different actors involved in light pollution, it is a slow process, and there are immediate threats to the quality of our night sky that must be addressed today by means of direct involvement. To this end, SOCHIAS and the OPCC are actively participating in an effort to coordinate several ongoing initiatives that can potentially have a strong impact on the preservation of Atacama's dark skies.

The first of these is a major national programme being pushed forward by the Ministry of Energy of Chile and the Chilean Agency for Energy Efficiency (AChEE in Spanish), which aims to replace 200,000 streetlights across the country for energy efficient LED lights during the next four years. The programme is expected to translate into a nationwide 20% decrease in power consumption associated with street lightning. Alarmingly, due to oversight and lack of coordination, the first incarnation of the programme did not consider the compliance with the recently published Decree Nº43 of the Ministry of Environment, which updated the light pollution control norms by limiting the amount of blue emission that new LED based outdoor lighting can produce. A swift reaction by the part of the OPCC, SOCHIAS and the astronomical observatories, together with a positive response from the Ministry of Foreign Relations and the agencies pushing forward the programme, now ensures that all the new luminaires installed in protected regions as part of this initiative will be compliant with the new standards.

Another experience that shows the need for constant oversight and coordination is an illumination project undertaken by the Chilean Ministry of Public Works on a stretch of the Panamerican Highway that crosses right in front of the La Silla and Las Campanas observatories. As part of a major upgrade to the highway and in an effort to comply with current safety standards for road illumination, all new intersections and a newly installed toll plaza were fitted with extremely bright lightning. Sadly this project was approved and executed right when the updated light pollution regulations were coming out, so it did not have to comply with the 'full cut off' restrictions nor with the maximum illumination levels contemplated in the new legislation.

Figure 3 shows two photographs taken from Las Campanas Observatory, more than 1,000 m above the level of the highway, before and after this upgrade. It is impressive to see the large amounts of light being emitted towards the sky from the highway, which, as also seen in Figure 1, is a major source of light pollution in the area. Not only is this project polluting the dark skies of one of the finest astronomical sites in the world, it also implies a major waste of resources since the new regulation requires all these luminaires to become compliant within a five year period, much shorter than the actual lifespan of these expensive systems.

A greater level of oversight and coordination could have prevented this problem. Thankfully, concerns raised by the OPCC after the project was executed have been positively received by the Ministry of Public Works. Today, there are plans for transforming this stretch of highway into a pilot project for testing Amber LED technology (one of the most astronomy-friendly types of LED illumination) on Chilean roads. If all goes well, such installations could become standard around major observatories in the country.

In summary, there is a lesson to be learned from these experiences. We cannot assume that private parties and government agencies will be aware of light pollution and will consider it as an important priority during their decision making processes. Therefore, constant vigilance by interested parties like observatories, scientific societies and environmental agencies is fundamental. In our experience, even minimal levels of communication, education and coordination can be successful at making people understand the problem of light pollution and see the overall benefits of avoiding it.

### Protecting our Heritage and Final Conclusions

The final and perhaps most important initiative that we are currently involved in, concerning the protection of our night sky, is an effort to recognize the Outstanding Universal Value that these marvellous sites in the Atacama Desert have for all of humanity. These unique places, where the best astronomical observatories in the world reside, are our 'windows to the universe'. They are the places from which humanity learns about the cosmos, unveils its origins and comes to understand its place in the world. The Chilean government, in close collaboration with SOCHIAS and the OPCC, are seeking the protection and recognition of the cultural and natural value of these sites at both national and international levels.

For this purpose, as members of a specially appointed working group created by the Chilean Ministry of Foreign Relations, we are working on preparing an application to seek the recognition by the Chilean government of the main observatory sites in the Atacama

Desert as Protected Areas. The goal is for this protection at the national level to be followed up by an application for nomination to the UNESCO World Heritage List. Such recognitions would provide enormous leverage to institutions like SOHCIAS and the OPCC to push forward the protection of these sites against light pollution, but there is a more fundamental impact that we expect from such designations. The Protected Area and UNESCO World Heritage site status will help us find a place for the night sky in the minds of millions of Chileans. We hope that the citizens of our country will start considering the beautiful clear and dark sky of the Atacama Desert as a fundamental part of their heritage. Something they should be proud of and something that their country as a nation must preserve for all humanity to enjoy for centuries to come.

It is this concept of preserving a unique and beautiful natural laboratory, which provides us with a window to the universe and to ourselves that guides our efforts. Chile as a nation has the responsibility to protect what nature has endowed us with and it is exciting to see how governments, industries, scientists and the public are starting to share this realisation. There is still a lot of work to be done, but we are on the right path to ensure that future generations will have access to the same dark sky that we have inherited from our ancestors.